\documentclass[twocolumn,aps,pre,floatfix,superscriptaddress]{revtex4}
\pdfoutput=1 %This is for the ArXiv submission only
\usepackage{amsmath,amssymb,eucal,graphicx,bm}

\usepackage[colorlinks=true, urlcolor=blue, anchorcolor=blue, citecolor=blue,filecolor=blue,linkcolor=blue,menucolor=blue,pagecolor=blue]{hyperref}

\begin{document}
%\title{Correlation Between Earlier Misfortune and Eternal Failure}
\title{Infinite System of Random Walkers: Winners and Losers}
\author{P. L. Krapivsky}
\affiliation{Department of Physics, Boston University, Boston, Massachusetts 02215, USA}

\begin{abstract}
We study an infinite system of particles initially occupying a half-line $y\leq 0$ and undergoing random walks on the entire line. The right-most particle is called a leader. Surprisingly, every particle except the original leader may never achieve  the leadership throughout the evolution. For the equidistant   initial configuration, the $k^{\text{th}}$ particle attains the leadership with probability $e^{-2} k^{-1} (\ln k)^{-1/2}$ when $k\gg 1$. This provides a quantitative measure of the correlation between earlier misfortune (represented by $k$) and eternal failure. We also show that the winner defined as the first walker overtaking the initial leader has label $k\gg 1$ with probability decaying as $\exp\!\left[-\tfrac{1}{2}(\ln k)^2\right]$. 
\end{abstract}
\maketitle

\section{Introduction}

We are fascinated and inspired by those who started at the bottom, yet made it to the top.  As a toy model accounting for the success or failure, we consider the system of non-interacting individuals evolving according to independent identically distributed random processes. The model does not attempt to mimic realistic social settings, yet emerging leadership characteristics are subtle, so the model illuminates potentially counter-intuitive features that may appear in more realistic situations. As random processes, we take independent and identical random walks.  In this case, the model is close to the models appearing in the realm of diffusion-controlled processes \cite{SR_book,R85,OTB89,book}, and it also mimics the search of a target by a swarm of diffusing particles (see \cite{Carlos} and references therein). 

Our model posits that every individual possesses an attribute (like popularity or wealth) varying according to a random process. We further assume that each attribute is a scalar quantity and postulate that all random processes are identical random walks.  The details of the random walk process are not important (it could be a Brownian motion, a continuous time random walk on the line or one-dimensional lattice, a discrete in space and time random walk). What is important is that random walks are assumed to be independent.  Achieving the leadership at a certain time(s) is classified as a success, failing to achieve the  leadership throughout the entire evolution is the failure. Our model is arguably the simplest social dynamics process allowing one to explore the leadership characteristics. 

The precise definition of the model is as follows. Denote by $y_j(t)$ the value of the attribute possessed by the $j^{\rm th}$ individual at time $t$. We shall think about $y_j(t)$ as the position of the $j^{\rm th}$ particle. Particles undergo identical random walks (or Brownian motions). We label particles in such a way that initially $y_1(0)>y_2(0)>\ldots$. By definition, the larger $y_j(t)$, the more successful the $j^{\rm th}$ individual is at time $t$. The $j^{\rm th}$ particle is called the leader at time $t$ if $y_j(t)>y_k(t)$ for all $k\ne j$. If a particle has never achieved the leadership throughout the evolution, $0<t<\infty$, we consider its history as an eternal failure. 

For two random walkers, the initial laggard surely catches the leader and the leap-frogging between the random walkers continues indefinitely. This is the basic feature of one-dimensional random walks \cite{SR_book,book,BM:book}. External failure is impossible for {\it finite} systems: Every random walker attains the leadership on infinitely many occasions and enjoys the leadership during an infinitely long time. The surprising result which we derive below is that eternal failure is {\em possible} in an infinite system. Furthermore, the eternal failure is the rule rather than exception --- although infinitely many random walkers achieve the leadership, the fraction of such walkers is infinitesimal. 

Denote by  $Q_k$ the probability that the $k^{\text{th}}$ particle has attained the leadership throughout its {\em entire} evolution history. In an infinite initially strictly ordered system, $y_1(0)>y_2(0)>\ldots$, the probabilities $Q_k$ constitute a strictly decreasing sequence:
\begin{equation}
\label{strict}
1\equiv Q_1>Q_2>Q_3>\cdots
\end{equation}
The first particle is initially the leader, so $Q_1=1$. The inequalities $Q_k<1$ for all $k\geq 2$ can be deduced, as we argue below, from well-known properties of Brownian motions such as the law of the iterated logarithm. The non-strict monotonicity, $Q_k\geq Q_{k+1}$ for all $k\geq 1$, is obvious. 
Proving that for an arbitrary strictly ordered initial configuration the probabilities $Q_k$ constitute a strictly decreasing sequence remains a challenge, although the validity of \eqref{strict} seems clear. 

Another general feature of the success probabilities $Q_k$ is the limiting behavior
\begin{equation}
\label{limit}
\lim_{k\to\infty}Q_k = 0
\end{equation}
This limiting behavior should certainly hold at least for non-pathological initial configurations spanning over the entire half-line. 

We mostly focus on the equidistant initial configuration. We argue that in this situation the probabilities $Q_k$ exhibit a peculiar asymptotic decay
\begin{equation}
\label{Qk:asymp}
Q_k \simeq e^{-2}  k^{-1} (\ln k)^{-1/2} \qquad\text{when}\quad k\gg 1
\end{equation}
This provides a quantitative version of \eqref{limit} in the specific case of the equidistant initial configuration. 
Equation \eqref{Qk:asymp} implies that among the first $N$ particles, the average number of those who have ever achieved the leadership exhibits an anomalously slow $2e^{-2}\sqrt{\ln N}$ growth in the $N\to \infty$ limit.  This asymptotic quantifies the assertions that (i) infinitely many random walkers achieve the leadership, and (ii) the fraction of such walkers is infinitesimal. 

We emphasize that despite the lack of interactions, the leadership characteristics are subtle and mostly unknown even for systems of a few random walkers (see \cite{Nied83,FG88,dbA88,Bramson91,KR96,KR99,laggard03,Cardy03,bk1,bk2,dbA18} and references therein). The infinite system is somewhat simpler and e.g. the inequalities $Q_k<1$ for $k\geq 2$ are easy to appreciate as we show below. This strict inequality may be summarized by an adage: {\em not a leader may never become a leader}. 

We now derive \eqref{Qk:asymp} for the equidistant initial configuration. We shall also look at the identity of the {\em winner} defined as the random walker which first overtakes the initial leader.

\section{Eternal Failure}
\label{sec:eternal}

Consider an infinite system of random walkers initially distributed uniformly with density $\rho$ on the half-line $y<0$ and undergoing identical Brownian motions with diffusion coefficient $D$. The initial density profile 
\begin{equation}
n(y,t=0) =
\begin{cases}
0       & y>0\\
\rho   & y\leq 0
\end{cases}
\end{equation}
evolves into 
\begin{equation}
\label{erfc:sol}
n(y,t) = \frac{\rho}{2}\,{\rm Erfc}\!\left(\frac{y}{\sqrt{4Dt}}\right)
\end{equation}
where ${\rm Erfc}(\eta)=\frac{2}{\sqrt{\pi}}\int_\eta^\infty du\,e^{-u^2}$ is the error function.

The average number of particles on the half-line $y\geq x$ is $\int_x^\infty dy\,n(y,t)$. Equating this number to unity gives an estimate of the position of the leader $z(t)$:
\begin{equation}
\label{leader:crit}
\int_{z(t)}^\infty dy\,n(y,t) \sim 1
\end{equation}
Plugging \eqref{erfc:sol} into \eqref{leader:crit} we obtain
\begin{equation}
\label{leader}
\rho z(t) \simeq \sqrt{2\tau \ln \tau}, \quad \tau=\rho^2 Dt
\end{equation}
when $\tau\to\infty$. The position of the leader is a random quantity that becomes more and more deterministic as $\tau\to\infty$, viz. the ratio of its standard deviation to the average vanishes as $\tau\to\infty$. More precisely:
\begin{equation}
\label{leader:determ}
\langle z\rangle \simeq \rho^{-1}\sqrt{2\tau \ln \tau}, \qquad 
\frac{\sqrt{\langle z^2\rangle - \langle z\rangle^2}}{\langle z\rangle}\sim \frac{1}{\ln \tau}
\end{equation}
The gap $g(t)$ between the leader and the following particle exhibits the same average growth as the standard deviation of the position of the leader:  
\begin{equation*}
\langle g(\tau) \rangle \sim \rho^{-1}\sqrt{\frac{\tau}{\ln \tau}}
\end{equation*}
This estimate can be extracted from the criterion
\begin{equation*}
\int_{z(t)-g(t)}^{z(t)} dy\,n(y,t) \sim 1
\end{equation*}

Therefore $z(t)$ is an asymptotically deterministic quantity growing (for $t\gg 1$) as $z=\rho^{-1}\sqrt{2\tau \ln \tau}$.  The above rather heuristic treatment can be made rigorous. A comprehensive description of the distribution of the position of the leader, the gap, etc. is presented, e.g., in Ref.~\cite{SS07}. 

To prove that $Q_k<1$ for $k\geq 2$ we employ the law of the iterated logarithm \cite{BM:book} describing moderate deviations of a random walk. Take a random walk and the boundary receding as $C\sqrt{4Dt\ln(\ln \tau)}$. The law of the iterated logarithm asserts that for $C<1$ and arbitrary $T$, the random walk surely crosses the boundary at some $t>T$;  for $C>1$, the probability of crossing at some $t>T$ decays to zero when $T\to\infty$. 

In our problem, the role of the deterministically receding boundary is played by the position of the leader. This position is a random quantity (as well as the identity of the leader). However, the position of the leader is asymptotically deterministic, see Eq.~\eqref{leader:determ}. Therefore we can use the law of the iterated logarithm. The position of the leader advances much faster than the threshold receding appearing in the law of of the iterated logarithm: $\sqrt{2\tau \ln \tau}\gg 2\sqrt{4\tau \ln(\ln \tau)}$. Hence the law of the iterated logarithm implies that $Q_k<1$ for $k\geq 2$. 

Take now the $k^{\text{th}}$ particle with $k\gg 1$. The probability 
$Q_k(t)$ that this particle becomes the leader during the time interval $(0, t)$ increases as
\begin{equation}
\label{Qkt}
\frac{dQ_k(t)}{dt} = -D\,\frac{\partial P_k(t,z)}{\partial z}
\end{equation}
where $P_k(t,z)$ is the probability density that the particle initially located at $y_k(0)= -\rho^{-1}k$ is at position $z(t)$ for the first time during time interval $(0,t)$. This probability density can be taken as the unperturbed Gaussian distribution
\begin{equation}
\label{Pzt}
P_k(t,z)=\frac{1}{\sqrt{4\pi Dt}}\,\exp\!\left\{- \frac{(z+\rho^{-1}k)^2}{4Dt}\right\} 
\end{equation}
The calculation scheme \eqref{Qkt}--\eqref{Pzt} is an approximation that should become asymptotically exact when $k\to \infty$. The ultimate probability is 
\begin{equation}
\label{Qk:inf}
Q_k \equiv Q_k(\infty) = \int_0^\infty dt \,\frac{dQ_k(t)}{dt}
\end{equation}
Using \eqref{Qkt}--\eqref{Pzt} and changing variables from $t$ to $Z=\rho z$ we transform \eqref{Qk:inf} to 
\begin{equation}
\label{Qk:int}
Q_k = \frac{1}{\sqrt{\pi}}\int_1^\infty dZ\,\sqrt{\ln Z}\,e^{F_k(Z)}
\end{equation}
where 
\begin{equation}
\label{Fk:def}
F_k(Z) = \ln\!\left(\frac{Z+k}{Z^2}\right) - \left(\frac{Z+k}{Z}\right)^2\ln Z
\end{equation}
Solving $\frac{dF_k(Z)}{dZ}=0$ we find that $F_k(Z)$ reaches the maximum at 
\begin{equation}
Z=k\zeta, \quad \zeta=\ln k + \ln(\ln k) + \ldots
\end{equation}
Near this maximum
\begin{equation}
\label{Fk:asymp}
F_k =  - 2\zeta -2 -\frac{(Z-k\zeta)^2}{(k\zeta)^2} + \ldots
\end{equation}
Plugging \eqref{Fk:asymp} into \eqref{Qk:int} and computing the Gaussian integral we arrive at the announced asymptotic \eqref{Qk:asymp}. 

The convergence to the asymptotic \eqref{Qk:asymp} is very slow since the sub-leading term differs from the leading term by the factor that vanishes as $\frac{\ln(\ln k)}{\ln k}$. Note that the quantities $Q_k$ do not depend on the initial density $\rho$ and the diffusion coefficient $D$. This is a general feature that can be appreciated on dimensional grounds: The probabilities $Q_k$ are dimensionless, while the quantities $\rho$ and $D$ have independent dimensions.

We now briefly comment on finite systems. The eternal failure becomes impossible \cite{impossible}, a particle which is not the initial leader would not attain the leadership during the time interval $(0,t)$ with probability decaying as $t^{-\psi_N}$ in the $t\to \infty$ limit. The amplitudes depend on the label $k=2,\ldots,N$  and details of the initial condition, while the persistence exponent $\psi_N$ depends only on the number of particles $N$. Two persistence exponents are known: $\psi_2 = \frac{1}{2}\,, \quad \psi_3 = \frac{3}{8}$. The exponent $\psi_2=1/2$ is the classical first passage exponent \cite{SR_book}. The exponent $\psi_3 = \frac{3}{8}$ can be derived by combining three independent one-dimensional Brownian motions into a single three-dimensional Brownian motion ${\bf y}=(y_1,y_2,y_3)$. The probability that ${\bf y}$ remains in the region where $y_3<y_2$ or $y_3<y_1$ is just the probability for a Brownian motion to remain in a wedge with opening angle $4\pi/3$. This probability is known to decay as  $t^{-\pi/2\theta}$ (see \cite{SR_book}) in the general case of the wedge with opening angle $\theta$, leading to  $\psi_3=3/8$ in the case of $\theta=4\pi/3$. 

The value of $\psi_4$ is known numerically \cite{laggard03,bk2,dbA18}, the most precise value $\psi_4\approx 0.3076360$ was recently found in Ref.~\cite{dbA18} through the mapping onto a related electrostatic problem. We finally note the asymptotic behavior 
\begin{equation*}
\label{psin} 
\psi_N= \frac{\ln N}{N}- \frac{\ln(\ln N)}{2N}+\ldots 
\end{equation*}
of the persistence exponent established in \cite{laggard03}.

\section{Winners}
\label{sec:winners}

Let us treat the initial leader as a target. This leader will be surely overtaken, and we seek the identity of the winner, namely the label $k$ of the random walker that first catches the initial leader, and also the time when this happens. Let $W_k$ be  the probability that the $k^\text{th}$ walker is the winner. Similarly to \eqref{strict} and \eqref{limit} we anticipate that these probabilities are strictly ordered
\begin{equation}
\label{strict-W}
1>W_2>W_3>\cdots
\end{equation}
and
\begin{equation}
\label{limit-W}
\lim_{k\to\infty}W_k = 0
\end{equation}
The inequalities $W_k<Q_k$ follow from the definitions of $W_k$ and $Q_k$. We also mention the obvious sum rule $\sum_{k\geq 2}W_k=1$. The challenge is to determine the asymptotic behavior of $W_k$. 

In the extreme situation when the initial leader is immobile, $y_1(t)=0$, the probabilities $W_k$ have been established in Ref.~\cite{RM}. Our problem is more complicated. Indeed, the natural idea of treating $x_k=y_1-y_k$ for $k\geq 2$ as coordinates of auxiliary particles does not  reduce our problem to the one studied in \cite{RM} because the auxiliary particles undergo correlated motions. This lack of correlations in the case of the stationary target allows to obtain  \cite{RM} exact predictions for $W_k$ for all $k\geq 2$. In the case of the diffusing target, it is not clear how to derive an explicit {\em exact} expression even for the simplest probability $W_2$. Similarly to external probabilities studied in Sect.~\ref{sec:eternal}  it is possible, however, to establish the asymptotic behavior of $W_k$. 

In our derivation, we use the probability $P(1,t)$ that the original leader maintained the lead throughout the time interval $(0,t)$. The asymptotic decay of this quantity is known  \cite{KR96,KR99}:
\begin{equation}
\label{P1t}
P(1,t) \asymp \exp\!\left[-\tfrac{1}{8}(\ln \tau)^2\right]
\end{equation}
(The notation $A(t)\asymp B(t)$ means the asymptotic equality of the logarithms: $\lim_{t\to\infty} \frac{\ln A(t)}{\ln B(t)}=1$.)

We now determine the asymptotic behavior of $W_k$. 
The particle with label $k\gg 1$ will be the winner if (i) the target was not caught for a long time; and (ii) at the moment when the $k^\text{th}$ particle overtakes the target, it is close to $z(t)$. The same argument as in deriving \eqref{Qkt}--\eqref{Qk:inf} together with Eq.~\eqref{P1t}, which turns into $\exp\!\left[-\tfrac{1}{2}(\ln Z)^2\right]$ in terms of $Z$, lead to an estimate 
\begin{equation}
\label{Wk:int}
W_k \asymp \int_1^\infty dZ\,\sqrt{\ln Z}\,e^{G_k(Z)}
\end{equation}
with 
\begin{equation}
\label{Gk:def}
G_k(Z) = F_k(Z) - \frac{1}{2}(\ln Z)^2
\end{equation}
and $F_k$ given by \eqref{Fk:def}. The quantity $G_k(Z)$ 
reaches the maximum at 
\begin{equation}
\label{ZC}
Z=k\big(\sqrt{3}+1\big) + \frac{Ck}{\ln k} + \ldots
\end{equation}
and the maximum value is 
\begin{equation}
\label{Gk:asymp}
G_k^\text{max} =   - \frac{1}{2}(\ln k)^2 + C_1\ln k + \ldots
\end{equation}
leading to
\begin{equation}
\label{Wk:asymp}
W_k \asymp \exp\!\left[-\tfrac{1}{2}(\ln k)^2\right]
\end{equation}
One can compute the amplitudes in \eqref{ZC}--\eqref{Gk:asymp} and also expand $G_k$ near the maximum and determine the Gaussian integral. This gives an algebraic $O(k^\mu)$ pre-factor to the asymptotic \eqref{Wk:asymp}. We do not write this pre-factor since the pre-factor in \eqref{P1t}  is unknown, and it probably also leads to an algebraic pre-factor. The decay law \eqref{Wk:asymp} is fast, namely faster than any power-law, but it is much slower than the  $\ln W_k\sim - k^{2/3}$ decay arising in the case of the stationary target \cite{RM}. 

If the winner has label $k\gg 1$, the time of the capture is found from $Z=\sqrt{2\tau\ln\tau}$ and \eqref{ZC} to give
\begin{equation}
\tau_k = \frac{k^2}{\ln k}\left(1+\frac{\sqrt{3}}{2}\right)
\end{equation}
This time is asymptotically deterministic. In the case of the stationary target the characteristic hitting time grows slower, viz. it scales as $k^{4/3}$, see \cite{RM}.

The search of an immobile target by a swarm of diffusive searches has been studied in numerous publications, see  \cite{BP77,ZKB83,T83,RK84,BZK84,BO87,Oshanin,BB03,KMR10,BMS13,MVK} and \cite{Carlos} for a review. The problem with a mobile target is much more challenging and exact treatments appear to be impossible, but the asymptotic behaviors are often essentially the same as in the case of the immobile target if the searches are uniformly distributed throughout the entire space. The mobility of the target is crucial in the inhomogeneous settings, particularly in one dimension with a one-sided distribution of searches. 

Let us briefly look at the situation when the diffusion coefficient of the target, $\mathcal{D}>0$, may differ from the diffusion coefficient $D$ of the searches. The probabilities $Q_k$ depend on the ratio $\delta=\mathcal{D}/D$, yet e.g. the asymptotic behavior \eqref{Qk:asymp} remains the same as long as $\delta>0$. The asymptotic behavior of the probabilities $W_k$ is slightly affected. Indeed, the generalization of the asymptotic \eqref{P1t} reads $P(1,t) \asymp \exp\!\left[-\tfrac{1}{8\delta}(\ln \tau)^2\right]$, see \cite{KR99}. The same analysis as before yields 
\begin{equation}
\label{Wk:Tk}
\begin{split}
W_k  &\asymp \exp\!\left[-\tfrac{1}{2\delta}(\ln k)^2\right] \\
\tau_k &= \frac{k^2}{\ln k}\left(\delta+\frac{\sqrt{\delta^2+2\delta}}{2}\right)^2
\end{split}
\end{equation}
Thus the ratio $\delta=\mathcal{D}/D$ only quantitatively affects the asymptotic behaviors of $W_k$ and $\tau_k$. Only when the ratio vanishes, that is, the target is immobile, the qualitatively different behavior emerges. 

\section{Concluding Remarks}
\label{sec:concl}

We have investigated the leadership characteristics of an infinite system of one-dimensional random walkers initially occupying the half-line. We have analyzed eternal properties quantifying the entire evolution history, specifically the probabilities $Q_k$ to ever achieve the leadership. These probabilities decay in a peculiar manner, Eq.~\eqref{Qk:asymp}. It is not clear how to derive an explicit {\em exact} expression even for the simplest such quantity $Q_2$. The justification of \eqref{strict}--\eqref{limit} for an arbitrary initial configuration, strictly ordered and spanning over the entire line, is another obvious challenge. 

Looking at the problem from a different angle, we have treated the initial leader as the target and other random walkers as the searches and examined the identity of the winner (the first random walker catching the target) and the time when this capture happens. Similarly to the situation with $Q_2$, explicit {\em exact} results seem impossible to derive already for $W_2$ and $\tau_2$. The asymptotic behaviors of $W_k$ and $\tau_k$ can be probed analytically, although the asymptotic \eqref{Wk:asymp} for $W_k$ gives only the dominant factor and hence it is significantly less precise than the asymptotic \eqref{Qk:asymp} for $Q_k$. We have also derived asymptotic results for $W_k$ and $\tau_k$ when the diffusivities of the target and searches differ,  Eq.~\eqref{Wk:Tk}. 

The apparent simplicity of the infinite system of non-interacting random walkers is misleading, the leadership characteristics are subtle and many basic questions remain unanswered. We have focused on the deterministic initial condition, namely the evolution starting from the equidistant configuration of random walkers. Different results emerge for annealed initial conditions, namely when we average over all initial conditions with uniform density. This is well-known in the case of the immobile target: The probability that the target has not been captured is $S(t)=\exp\!\big[-\pi^{-1/2}\sqrt{\tau}\big]$ in the annealed case, while in the equidistant case the decay is faster, $S(t)\asymp \exp\!\big[-\Lambda \sqrt{\tau}\big]$ with $\Lambda > \pi^{-1/2}$, see \cite{RM}. The asymptotic behaviors of $Q_k$ and $W_k$ in the annealed case are unknown.  

The  problem with stationary target has been reformulated \cite{MVK} using large deviation techniques, more precisely a macroscopic fluctuation theory (MFT), see \cite{MFTreview} for a review of the MFT. This reformulation reproduces previously known results for the survival probability of the immobile target and additionally predicts the most probable evolution of the density of random walkers resulting in the very long survival of the target. It would be interesting to extend the analysis \cite{MVK} to the case of the mobile target survived during a long time. 

In this work, we have studied eternal leadership characteristics. Understanding the temporal behaviors is an obvious direction for future work. For instance, one would like to establish the average total number of different random walks achieving the leadership during the long time interval $(0,t)$. The relaxation characteristics, e.g., the long time decay of $Q_k(\infty)-Q_k(t)$, are also interesting.

\bigskip\noindent
I am grateful to Eli Ben-Naim, Baruch Meerson and Sid Redner for discussions.


\begin{thebibliography}{99}
\bibitem{impossible}
      Eternal failure is impossible in the most general situation of arbitrary initial positions and arbitrary diffusion 
      constants $D_1, \ldots, D_N$.
\end{thebibliography}
\end{document}